\documentclass[journal]{IEEEtran}

\usepackage{times}
\usepackage{graphicx} 
\usepackage{subfigure} 

\usepackage{algorithm}
\usepackage{algorithmic}
\usepackage{setspace}

\usepackage{helvet}
\usepackage{courier}
\usepackage{hyperref}

\usepackage{makecell}
\usepackage{graphicx}
\usepackage{subfigure}
\usepackage{color}
\usepackage{amsfonts}
\usepackage{amsmath}
\usepackage{amssymb}

\usepackage{multirow}
\usepackage{booktabs}
\usepackage{cleveref}

\DeclareMathAlphabet\mathbfcal{OMS}{cmsy}{b}{n}

\def\0{{\bf 0}}
\def\1{{\bf 1}}

\usepackage{ntheorem}

\newtheorem*{*thm}{Theorem}

\newtheorem*{*lemma}{Lemma}

\def\ie{\mbox{\textit{i.e.}}}

\usepackage{subfigure}
\usepackage{multirow}
\usepackage{url}
\usepackage{cite}
\usepackage{hyperref}
\usepackage{microtype}

\begin{document}

\title{Perception- and Fidelity-aware Reduced-Reference Super-Resolution Image Quality Assessment}

	\author{Xinying Lin, Xuyang Liu, Hong Yang, Xiaohai He, \textit{Member, IEEE}, and Honggang Chen, \textit{Member, IEEE}
	\IEEEcompsocitemizethanks{
	\IEEEcompsocthanksitem{This work was supported in part by the National Natural Science Foundation of China under Grant 62001316, in part by Sichuan Science and Technology Program under Grant 2024YFHZ0212, in part by the Open Foundation of Yunnan Key Laboratory of Software Engineering under Grant 2023SE206, in part by the Research Fund of Guangxi Key Lab of Multi-source Information Mining\& Security under Grant MIMS22-14, and in part by the Fundamental Research Funds for the Central Universities under Grant SCU2023D062 and under Grant 2022CDSN-15-SCU.  \textit{(Corresponding author: Honggang Chen.)}

    Xinying Lin is with the College of Electronics and Information Engineering, Sichuan University, Chengdu 610065, China, and also with Guangxi Key Lab of Multi-source Information Mining Security, Guangxi Normal University, Guilin 541004, China (email: linxinying@stu.scu.edu.cn).

    Xuyang Liu, Hong Yang, and Xiaohai He are with the College of Electronics and Information Engineering, Sichuan University, Chengdu 610065, China (email: liuxuyang@stu.scu.edu.cn; yhscu@scu.edu.cn; hxh@scu.edu.cn).
    
    Honggang Chen is with the College of Electronics and Information Engineering, Sichuan University, Chengdu 610065, China, and also with the Yunnan Key Laboratory of Software Engineering, Yunnan University, Kunming 650600, China (e-mail: honggang\_chen@scu.edu.cn).}
	}
    }

\maketitle
\begin{abstract}
With the advent of image super-resolution (SR) algorithms, how to evaluate the quality of generated SR images has become an urgent task. Although full-reference methods perform well in SR image quality assessment (SR-IQA), their reliance on high-resolution (HR) images limits their practical applicability. Leveraging available reconstruction information as much as possible for SR-IQA, such as low-resolution (LR) images and the scale factors, is a promising way to enhance assessment performance for SR-IQA without HR for reference. In this paper, we attempt to evaluate the perceptual quality and reconstruction fidelity of SR images considering LR images and scale factors. Specifically, we propose a novel dual-branch reduced-reference SR-IQA network, \ie, Perception- and Fidelity-aware SR-IQA (PFIQA). The perception-aware branch evaluates the perceptual quality of SR images by leveraging the merits of global modeling of Vision Transformer (ViT) and local relation of ResNet, and incorporating the scale factor to enable comprehensive visual perception. Meanwhile, the fidelity-aware branch assesses the reconstruction fidelity between LR and SR images through their visual perception. The combination of the two branches substantially aligns with the human visual system, enabling a comprehensive SR image evaluation.
Experimental results indicate that our PFIQA outperforms current state-of-the-art models across three widely-used SR-IQA benchmarks. Notably, PFIQA excels in assessing the quality of real-world SR images.

\end{abstract}


\IEEEpeerreviewmaketitle

\section{Introduction}

\IEEEPARstart{I}{mage} super-resolution (SR) technology aims to produce more detailed high-resolution (HR) images from the given low-resolution (LR) images. It has been widely used in various fields, such as security and surveillance\cite{gao2023jdsr}, medical imaging\cite{georgescu2023multimodal}, and remote sensing imaging\cite{xiao2023ediffsr,2024shinRemote}. Given the remarkable advancements in recent research on SR algorithms, including blind SR\cite{li2023learning,Neshatavar_2024_WACV}, lightweight SR models\cite{wan2020lightweight,esmaeilzehi2021srnmsm,wang2024osffnet},  arbitrary scale SR\cite{zhao2024tmm}, and multimodal SR\cite{ZHOU_2024_ICASSP, Noguchi_2024_WACV}, it has become imperative to evaluate the quality of the generated SR images. This evaluation is crucial for facilitating comparative analysios of reconstruction performance across various SR models and guiding the development of SR algorithms.

Numerous methods have been proposed for image \mbox{quality} assessment (IQA) \cite{liao2023full,hu2023blind,mittal2012no,mittal2012niqe,2022maniqa}, which can be categorized as subjective or objective. While subjective ones are more \mbox{reliable}, they are impractical due to high costs and external factors. Hence, objective IQA methods that align with subjective \mbox{evaluations} are currently a focus of research. IQT\cite{cheon2021perceptual}, MANIQA\cite{2022maniqa} and AHIQ\cite{lao2022attentions} are recently proposed IQA methods that achieve excellent subjective consistency.
While generic IQA methods have shown satisfactory results, they often neglect the specific characteristics of SR, making them unsuitable to directly apply to SR images. 
SR algorithms aim to recover detailed information from LR images, so SR-IQA needs to not only focus on the visual quality of the image, but also consider the consistency with the LR image, a factor that generic IQA methods overlook. 
At the same time, SR images often exhibit mixed degradations, such as blurring, ringing, and aliasing artifacts, which are not effectively addressed by current generic IQA methods.

Recently, there has been an increase in the development of SR-IQA methods\cite{zhou2024database}.
Depending on the availability of lossless reference images, SR-IQA methods can be categorized into full-reference\cite{zhang2023perception}, reduced-reference\cite{zhao2021learning,Noguchi_2024_WACV}, and no-reference methods\cite{fu2023scale, li2022c}. 
For all three SR-IQA paradigms, evaluating the \textbf{\textit{perceptual quality}} is of paramount importance, as it directly relates to human perceptual judgments and assessment of the given SR images\cite{zhang2022spqe,zhang2023perception}. Moreover, \textbf{\textit{reconstruction fidelity}}\cite{huang2023refsr,luo2024skipdiff} is also highly important since it reflects how faithfully the SR image represents the reference image details and content. Only reference-based SR-IQA methods can effectively consider reconstruction fidelity by comparing the SR image against the reference image.
However, in practical scenarios where HR images are unavailable, the application of full-reference SR-IQA methods becomes challenging. Fortunately, the SR task inherently possesses available reference or auxiliary information, such as the paired LR image and the scale factor. These cues have significant reference value for evaluating SR images\cite{zhang2023perception}, but are not directly utilized by most SR-IQA methods\cite{li2022c,fu2023scale}. Recently, a few reduced-reference SR-IQA methods \cite{zhao2021learning,zhang2023perception,zhou2023RR-IQA} have adopted CNN-based networks to get feature maps of LR and SR images, and simply calculated the similarity between the feature maps and utilized multi-layer perceptron to regress the quality scores.
The entire process can be seen as a pixel-by-pixel comparison between the LR and SR images, which yet neglects the spatial coherence between the LR and SR images, resulting in an insufficient emphasis on reconstruction fidelity.

To address these issues, we present a novel reduced-reference \underline{\textbf{P}}erception- and \underline{\textbf{F}}idelity-aware SR-\underline{\textbf{IQA}} (\textbf{PFIQA}), which integrates LR images and scale factors as prior \mbox{knowledge} to assist in SR-IQA. Specifically, PFIQA consists of two assessment branches: the Perception-aware Assessment Branch (PA Branch) for  evaluating \textbf{\textit{perceptual quality}} and the Fidelity-aware Assessment Branch (FA Branch) for assessing \mbox{\textbf{\textit{reconstruction fidelity}}}. 
Given that each patch holds unique visual details, we design a patch scoring module for each branch and a patch weighting module to assign varying degrees of attention to each patch, thus achieving fine-grained patch-wise prediction for SR-IQA. The outputs of the two branches are combined through a sum to provide a comprehensive evaluation of the SR images. Based on this delicate design, PFIQA can enhance the consistency between the network's assessment results and human visual system (HVS).

Our main contributions can be summarized as follows:
\begin{itemize}
    \item We introduce PFIQA, a novel dual-branch reduced-reference SR-IQA network that comprehensively assesses the perceptual quality and reconstruction fidelity of SR images without requiring any reference HR images.
    
    \item Our proposed PFIQA takes pairs of SR and LR images as input, leveraging the merits of global modeling of ViT and local relation of ResNet to enable comprehensive visual perception, and incorporates the scale factor to effectively align with HVS.
    
    \item Extensive experiments on three widely-used SR-IQA benchmarks demonstrate that PFIQA shows superior performance compared with other SR-IQA methods.
\end{itemize}

\section{Related work}
\subsection{Image Quality Assessment}
We briefly review generic image quality assessment (IQA) methods, which can be broadly categorized into full-reference IQA (FR-IQA), no-reference IQA (NR-IQA) and reduced-reference IQA (RR-IQA). 

For FR-IQA, the most widely used metrics are PSNR and SSIM\cite{wang2004image}. However, despite their simplicity and ease of optimization, these metrics correlate poorly with HVS\cite{lao2022attentions}. To address the limitations of conventional IQA methods, various learning-based FR-IQA approaches have been recently proposed\cite{cheon2021perceptual,lao2022attentions}. 
IQT\cite{cheon2021perceptual} utilizes an encoder-decoder transformer with trainable quality embeddings for superior performance. AHIQ\cite{lao2022attentions} predicts image quality at the patch level, excelling in GAN-generated distortions.
For NR-IQA, current methods emphasize generality and can be further categorized into two types: natural scene statistics (NSS)-based metrics\cite{mittal2012niqe,mittal2012no} and learning-based metrics\cite{kang2014convolutional,2022maniqa}. NSS-based NR-IQA methods require manual extraction of image features, and these hand-crafted feature representations are often not effective. Subsequently, learning-based methods have demonstrated superior performance. CNNIQA\cite{kang2014convolutional} introduces a pioneering CNN for IQA, directly learning from image pixels without the need for hand-crafted features. MANIQA\cite{2022maniqa} proposes a multi-dimensional attention mechanism for interactions in both channel and spatial domains, employing ViT and novel modules to enhance global and local interactions. For RR-IQA, CKDN\cite{zheng2021learning} learns a reference space from degraded images to extract knowledge from pristine-quality images.

Although current IQA methods demonstrate satisfactory results, they often fail to account for the specific characteristics of SR images, making them inadequate for direct application in SR-IQA. SR images typically exhibit various mixed degradations, including blurring, ringing, and aliasing artifacts, which current generic IQA methods are not effective in addressing these issues. Therefore, there is a pressing need for specialized SR-IQA methods that take into consideration the unique degradation patterns and characteristics of SR images.

\subsection{Super-Resolution Image Quality Assessment}

SR-IQA approaches can be divided into two main categories: hand-crafted feature-based methods\cite{yeganeh2015objective,ma2017nrqm,zhou2019visual,zhou2022quality} and learning-based methods\cite{zhao2021learning,li2022c,fu2023scale,zhang2023perception}.

The hand-crafted feature-based methods perform regression to predict the quality scores of SR images. Yeganeh et al.\cite{yeganeh2015objective}  employ three statistical features for reduced-reference SR-IQA, including sub-image frequency energy falloff statistics, sub-image local dominant orientation statistics, and spatial continuity statistics. NRQM\cite{ma2017nrqm}adopts three types of low-level statistical features in both spatial and frequency domains to quantify SR artifacts for no-reference SR-IQA. SIS\cite{zhou2019visual} leverages structure-texture decomposition to evaluate the visual quality of SR images. 
However, these types of methods have shown limited effectiveness in practice.

For learning-based methods, the goal is to automatically map images to their corresponding quality scores. Li et al.\cite{li2022c} introduce C$^{2}$MT, a class-aware multi-task transformer that uses supervised contrastive learning and active learning. PCST\cite{zhang2023perception} addresses the challenge of balancing referenced and no-reference scores by integrating multi-scale and saliency information into deep learning modules. However, in real-world scenarios where HR images are unavailable, full-reference SR-IQA methods become impractical. DISQ\cite{zhao2021learning}, inspired by the VGG network, employs a dual-stream deep neural network architecture by utilizing LR images, which is considered a reduced-reference SR-IQA method. However, in addition to LR images, SR methods can also leverage certain priors, such as the scale factor. Fu et al.\cite{fu2023scale} present a scale-guided hypernetwork framework (SGH) for no-reference SR-IQA, demonstrating improved performance over existing IQA metrics through scale-adaptive evaluation.
While these cues are valuable for SR-IQA, most SR-IQA methods do not directly utilize them\cite{li2022c,fu2023scale}. Concurrently, some recent reduced-reference approaches\cite{zhao2021learning,zhang2023perception,zhou2023RR-IQA} use CNNs to extract and compare LR and SR image features, then employ MLPs for quality scoring. However, this pixel-level comparison fails to account for spatial coherence, undermining the assessment of reconstruction fidelity.
In this work, our proposed PFIQA utilizes LR images and scale factors as priors, combines ViT and CNN for comprehensive feature extraction, and adopts fine-grained patch-wise prediction for SR-IQA to assess both the perceptual quality and reconstruction fidelity of SR images.
\begin{figure*}[!ht]
\centering

\includegraphics[width=1.01\textwidth]{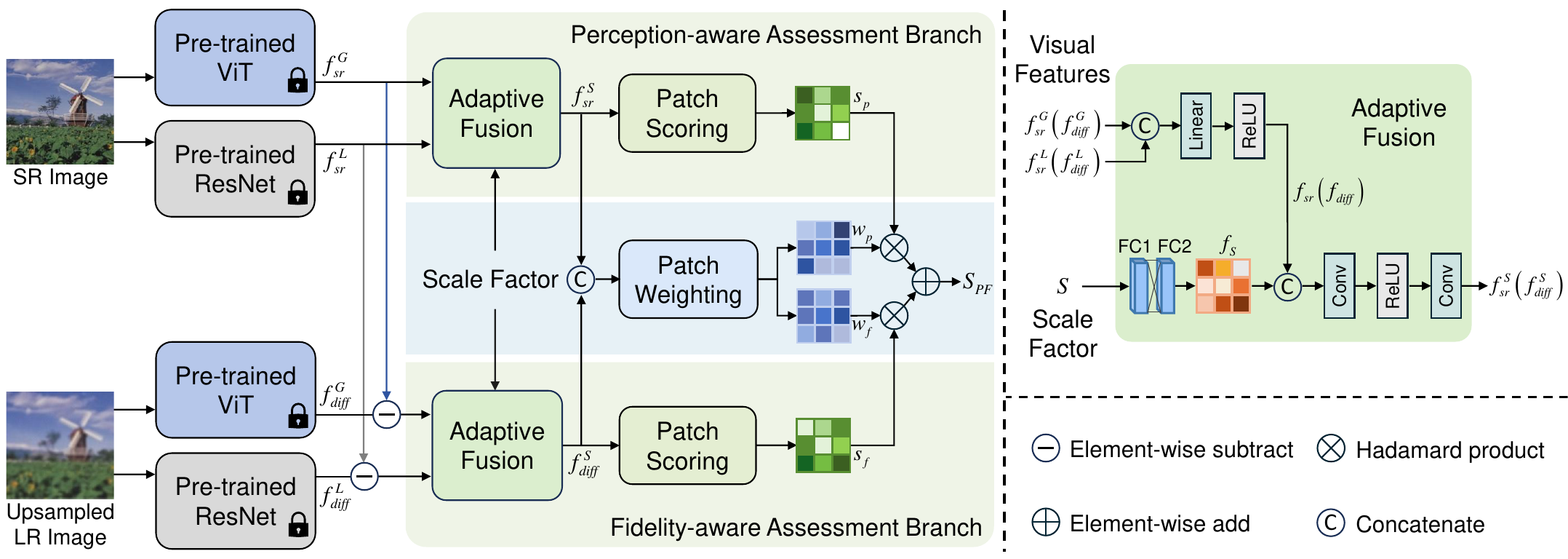}
\caption{{\textbf{Overview of proposed PFIQA method.} 
This framework consists of Perception-aware Assessment Branch (PA Branch) and Fidelity-aware Assessment Branch (FA Branch) for SR-IQA.}}
\vspace{-4mm}
\label{Fig:overview}
\end{figure*}

\section{Methodology}

As shown in Fig. \ref{Fig:overview}, our proposed PFIQA is composed of two assessment branches, underging three key phases: (1) \textbf{\textit{feature extraction}} to acquire global and local visual features; (2) \textbf{\textit{adaptive fusion}} of visual features and the scale factor to obtain fine-grained representations conducive for SR-IQA; (3) \textbf{\textit{patch-weighted quality regression}} to produce a perception-aware score and a fidelity-aware score and perform a sum of them to obtain the final quality score. 

\subsection{Feature Extraction}
\label{sec:FE}
The first phase of PFIQA is feature extraction, which consists of pre-trained Vision Transformer (ViT)\cite{dosovitskiy2020image} and ResNet \cite{he2016deep}. 
Understanding the broader context of an image is crucial for the IQA task as it provides essential information about the overall structure and content. ViT excels in this aspect by primarily focusing on extracting global visual features. Its self-attention mechanism effectively captures long-range dependencies and encodes images into comprehensive global feature representations, which are valuable for SR-IQA.
In addition to global features, paying attention to fine-grained details is equally important, especially in SR-IQA where humans tend to focus on intricate elements. Incorporating local visual information alongside global features can significantly enhance the accuracy and robustness of SR-IQA methods. Inspired by previous work \cite{saha2023re,chen2024topiq}, we augment ViT with ResNet to better capture local visual features, thereby enriching the model's ability to comprehensively represent images.

Specifically, a pair of LR and SR images are fed into ViT and ResNet, and then we take out the ViT feature maps of the selected stages and ResNet feature maps of different scales. Due to the similarity in the feature extraction process between SR and LR images, we illustrate the process taking the SR image as an example. For ViT, we utilize the outputs from five stages within the ViT backbone. 
The output feature from each block $f_{sr}\in \mathbb{R}^{p\times p \times c}$, where $c=768$, $p=28$, is concatenated into $f_{sr}^{G}\in \mathbb{R}^{p\times p \times 5c}$, and then reshaped into $f_{sr}^{G^{\prime}}
\in \mathbb{R}^{5c \times p \times p}$. For ResNet, we extract multi-scale features from four different stages of ResNet, and then interpolate these features to resize them to the same dimensions. After that, we concatenate them along the channel dimension and reshape them to obtain $f_{sr}^{L^{\prime}}
\in \mathbb{R}^{3840 \times p \times p}$. We apply a convolutional dimensionality reduction to get the final output global features $f_{sr}^{G}
\in \mathbb{R}^{256 \times p \times p}$ and local features $f_{sr}^{L}
\in \mathbb{R}^{256 \times p \times p}$. 
Similarly, the paired LR image undergoes the same process, resulting in the output of global features $f_{lr}^{G}$ and local features $f_{lr}^{L}$.

For PA branch, the target is to evaluate the perceptual quality of the SR image itself. Therefore, we utilize the extracted feature of the SR images $f_{sr}^{G}$ and $f_{sr}^{L}$ as input to PA branch.
In contrast, for FA branch, in order to measure the fidelity between the SR image and its paired LR image, we take into consideration that LR and SR images exhibit differences in the feature space. 
After obtaining the global and local visual features extracted respectively by ViT and ResNet, we represent the features $f_{diff}^{L}=f_{sr}^{L} - f_{lr}^{L}$ and $f_{diff}^{G}=f_{sr}^{G} - f_{lr}^{G}$ as input to FA Branch. These features are represented as the difference between the global and local features to capture the distinction between the SR image and its paired LR image.

\subsection{Adaptive Fusion}
To effectively combine the extracted visual features for the two branches, as well as auxiliary information from the scale factor for SR-IQA, we employ Adaptive Fusion Modules (AFMs). For each branch, the global and local features along with the scale factor, are utilized as inputs to the AFM. Since the processing by AFM in two branches is similar, with the only difference lying in their respective inputs, we provide a detailed description of the AFM in PA branch as an example.  As shown in Fig. \ref{Fig:overview} (right), we first adaptively fuse global and local visual features, and then incorporate the scale factor information.

\begin{table*}[t]
	\centering
	\renewcommand\arraystretch{1.1}
	\caption{An overview of widely-used SR-IQA benchmarks.}
 	\begin{tabular}{m{2.0cm}<{\centering}m{1.6cm}<{\centering}m{1.6cm}<{\centering}m{1.7cm}<{\centering}m{1.8cm}<{\centering}m{1.7cm}<{\centering}m{1.7cm}<{\centering}}
		\hline
		Datasets & LR Images&  SR Images & Number of SR methods& Scaling factors & Labels & Synthetic / Realistic \\
		\hline 
	   QADS\cite{zhou2019visual} & 60 & 980 & 21 & \{$\times$2,$\times$3,$\times$4\} & MOS &Synthetic\\ 		\hline 
        WIND\cite{yeganeh2015objective} & 13 & 312 & 8 & \{$\times$2,$\times$4,$\times$8\} & Rank &Synthetic\\ 		\hline 
		RealSRQ\cite{jiang2022single} & 180  & 1620 & 10 &  \{$\times$2,$\times$3,$\times$4\} & MOS   &Realistic\\

		\hline
	\end{tabular}
    \label{tab:dataset}
\end{table*}

\textbf{Global and Local Visual Features.}
The AFM is capable of learning the importance of features from global and local features and adaptively assigning appropriate weights to them, thereby obtaining comprehensive visual features for SR-IQA. 

For PA branch, global features ($f_{sr}^{G}
$) and local features ($f_{sr}^{L}
$) are first fed into the AFM and then concatenated along a new dimension.
Then, to effectively learn the weights of global and local features to fuse the two features adaptively, we employ a fusion operation implemented by a fully-connected layer followed by a ReLU activation function, which yields the features $f_{sr}$. Similarly, for FA branch, we fuse $f_{diff}^{G}$ and $f_{diff}^{L}$ to obtain $f_{diff}$.

\textbf{Scale Factor.}
The scale factor has a statistically significant influence on the subjective quality scores of SR images \cite{fu2023scale}, suggesting its potential as a valuable indicator for assessing the quality of SR images and guiding SR-IQA. Therefore, we input the vector representing the scale factor $S$ into the AFM to generate SR-IQA-related features. First, the scale factor is passed through two fully-connected layers and then reshaped to obtain the scale factor feature $f_S$. 

Subsequently, for PA branch, the scale factor feature $f_S$ is concatenated along the channel dimension with the features $f_{sr}$. The concatenated features then go through a $3\times3$ convolution, a ReLU activation, and another $3\times3$ convolution to obtain the perceptual features $f_{sr}^S$. Similarly, for FA branch, the differential features $f_{diff}^S$ are obtained.

\subsection{Patch-Weighted Quality Regression}

\label{subsec:pooling}
The final phase of PFIQA is patch-weighted quality \mbox{regression}, which comprises two patch scoring modules and a patch weighting module to generate the score $S_{PF}$ of the input SR image. 
Since each pixel in the deep feature map corresponds to a distinct patch of the input image and encapsulates abundant information, the spatial dimension's information is crucial. To capture the relationships between image patches, we predict scores for each pixel in the feature maps of the two branches and calculate the attention maps for each corresponding score. We obtain the scores for the two branches separately by performing a weighted sum of the individual scores. The weighted summation is employed to capture the significance of each region, simulating the behavior of the HVS. Finally, we add the scores from these two complementary branches to obtain the final predicted score.

Specifically, patch scoring module employs perceptual features $f_{sr}^S$ and differential features $f_{diff}^S$ to evaluate the perception and fidelity of the SR image in a patch-wise manner. Taking the PA branch as an example, the visual features $f_{sr}^S$ pass through a $3\times3$ convolution, followed by a ReLU activation, and another $3\times3$ convolution to produce the perception-aware score map $s_p$. Similarly, for the FA branch, the process yields the fidelity-aware score map $s_f$.

In parallel, the perceptual features $f_{sr}^S$ and differential features $f_{diff}^S$ are concatenated and fed into a patch weighting module, which computes the perception-aware score weight map $w_p$ and the fidelity-aware score weight map $w_f$ for each image patch. This process can be represented as:
\begin{equation}
\resizebox{0.9\hsize}{!}{$w = Sigmoid(Conv1(ReLU(Conv3(Concat(f_{sr}^S ,f_{diff}^S)))))$},
\end{equation}
where $Conv1$ and $Conv3$ represent $1\times1$ and $3\times3$ convolution, the score weight map $w$ is a two-channel tensor, with the first channel representing the perception-aware score weight map $w_p$ and the second channel representing the fidelity-aware score weight map $w_f$, \ie, $w = [w_p, w_f]$. 

Finally, we utilize the two score weight maps to perform weighted summation on the perception-aware score map and fidelity-aware score map and then add them to obtain the final predicted quality score $S_{PF}$. This can be represented as:
\begin{equation}
S_{PF} = \frac{s_{p} * w_{p}}{\sum w_{p}} + \frac{s_{f} * w_{f}}{\sum w_{f}},
\end{equation}
where $*$ means Hadamard product. 
\section{Experiments}
\label{exp_set}

\begin{table*}[t!]
\centering

\caption{Performance comparison on three benchmark datasets. We highlight the \textbf{best} and the \underline{second} results. Algorithms marked with asterisks (*) have been reproduced, while unmarked ones use results from the original paper. }
\begin{tabular}{c|c|c|cc|cc|cc|cc}

\toprule

\multirow{2}{*}{Type} & \multirow{2}{*}{Reference} & \multirow{2}{*}{Methods} & \multicolumn{2}{c|}{QADS\cite{zhou2019visual}} & \multicolumn{2}{c|}{WIND\cite{yeganeh2015objective}} & \multicolumn{2}{c|}{RealSRQ\cite{jiang2022single}} & \multicolumn{2}{c}{Average} \\

& & & PLCC & SRCC & PLCC & SRCC & PLCC & SRCC & PLCC & SRCC \\

\midrule
\multirow{7}{*}{Generic IQA} & 
HR&PSNR*& 0.3024 & 0.2949 &0.7455& 0.7431 & 0.0812& 0.0951&0.3764 &0.3777 \\
&HR&SSIM*\cite{wang2004image}&0.4862& 0.4833 &0.2899  & 0.4216  &0.1062 & 0.1319 &0.2941 & 0.3456\\

&$\times$&NIQE*\cite{mittal2012niqe}&0.3043 & 0.3478&0.2393 & 0.3195&0.0268 &0.0981& 0.1901&0.2551\\

& $\times$&BRISQUE*\cite{mittal2012no}&0.5282 & 0.5449&0.7475 & 0.7552&0.0515 & 0.0076&0.4424&0.4359\\

&$\times$&CNNIQA*\cite{kang2014convolutional}&0.8791 & 0.8742&0.9328 & 0.8902&0.6711 & 0.6671&0.8277&0.8105\\

&$\times$ & MANIQA*\cite{2022maniqa} & 0.9804 & 0.9761  & 0.9771 & 0.9321 & 0.8631 & 0.7812 & 0.9402 & 0.8964 \\

& HR & AHIQ*\cite{lao2022attentions} &  \underline{0.9815} &\underline{0.9805} & 0.9784 & 0.9522 & \underline{0.8786} & \underline{0.7882}& \underline{0.9461} & \underline{0.9069}  \\

\midrule
&HR&SIS \cite{zhou2019visual}&0.9137 & 0.9132 &0.8913 & 0.8777& -& -& -&-\\

\multirow{8}{*}{SR-IQA} &$\times$&NRQM*\cite{ma2017nrqm}&0.6867 & 0.6898&0.6869 & 0.6201&0.1451 & 0.0042&0.5062&0.4380\\

&$\times$&C$^{2}$MT \cite{li2022c}&0.9719 & 0.9690&- & -& 0.7184 & 0.7043& - & -\\

&$\times$&SRIF \cite{zhou2022quality}&0.9174 & 0.9163&0.9525 & 0.9157& -& -&-&-\\

&$\times$&SGH*\cite{fu2023scale}&0.9362 & 0.9475&0.9733 & \underline{0.9563}& 0.6847 & 0.7069&0.8647&0.8702\\

&LR&DISQ*\cite{zhao2021learning}&0.9127 & 0.9110&\underline{0.9785} & 0.9333&0.7885 & 0.7175&0.8932 &0.8539\\
&LR&PSCT\cite{zhang2023perception}&0.9620 & 0.9600 & 0.9596 & 0.9290   &- & -&-&-\\

\cmidrule{2-11}
&LR & PFIQA (Ours) & \textbf{0.9830} & \textbf{0.9815} & \textbf{0.9827} & \textbf{0.9637} & \textbf{0.9269} & \textbf{0.8597} & \textbf{0.9642} & \textbf{0.9350} \\

\bottomrule
\end{tabular}
\vspace{-4mm}
\label{tab:1}
\end{table*}

\begin{figure*}[!ht]
\centering
\includegraphics[width=1.00\textwidth]{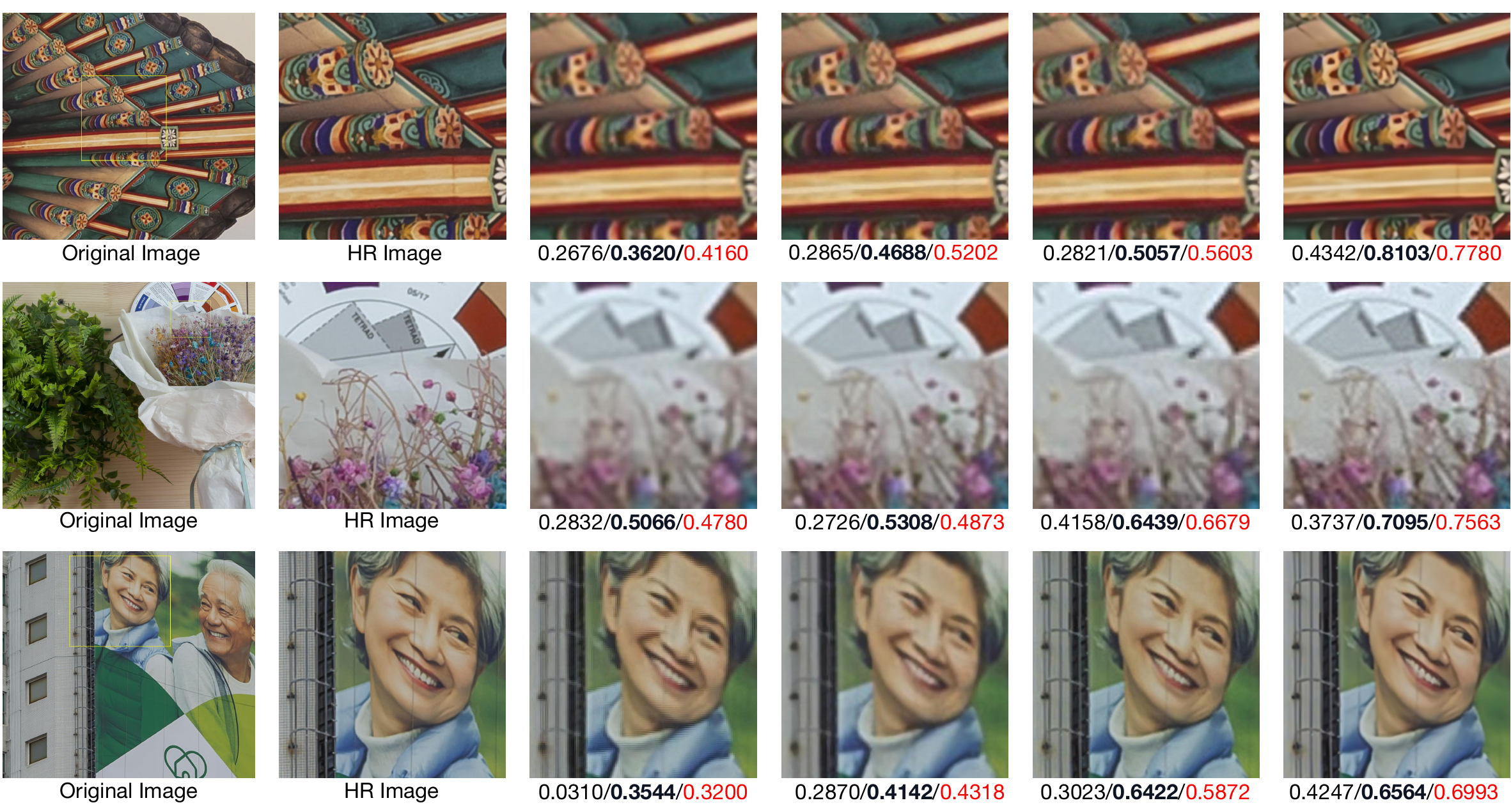}
\caption{Examples of predicted scores on RealSRQ\cite{jiang2022single} include DISQ\cite{zhao2021learning} (Previous SR-IQA SOTA) / \textbf{PFIQA (Ours)} / \textcolor{red}{MOS}. We consistently crop and enlarge each image for better visualization. The images are arranged in order of increasing image quality, with corresponding MOS values progressively increasing.}
\vspace{-4mm}
\label{Fig:score}
\end{figure*}

\subsection{Experimental Settings}
\textbf{Datasets and Evaluation Metrics.}
The evaluations are implemented on three datasets which are commonly used in the research of SR-IQA, including QADS\cite{zhou2019visual}, WIND\cite{yeganeh2015objective} and RealSRQ\cite{jiang2022single}. Table \ref{tab:dataset} describes the details of the above datasets. We split the datasets randomly into training and test sets at an 8:2 ratio, repeat the partitioning and evaluation process 5 times for fair comparison, and report the average results as the final performance. We utilize two of the most commonly used metrics, including Pearson Linear Correlation Coefficient (PLCC) and Spearman Rank-order Correlation Coefficient (SRCC). PLCC assesses the linear correlation between  mean opinion scores (MOSs) and predicted quality scores. It is used to measure the accuracy of IQA algorithm predictions, while SRCC measures the monotonicity of IQA algorithm predictions. Higher values of these two metrics indicate better performance of the IQA method.

\textbf{Implementation Details.}
We use ViT-B/8 \cite{dosovitskiy2020image} and ResNet-50 \cite{he2016deep} models pre-trained on ImageNet \cite{ILSVRC15}, which are not updated during training. We bilinearly interpolate the LR image to achieve the same resolution as the SR image. For model training, we normalize all input images and randomly crop them into a size of $224\times224$. Additionally, a random horizontal flip rotation is applied to augment the training data. To optimize the model, we utilize L2 Loss between the predicted score and the MOS, and the batch size is 4. During the test phase, we crop the four corners and the center of the original image, and the final score is the average of the scores from these cropped sections. We train PFIQA on a single NVIDIA GeForce RTX3090 GPU. The training process utilizes the AdamW optimizer,  setting the maximum number of training epochs to 200.

\subsection{Main Results}

\begin{figure*}[!ht]
\centering
\includegraphics[width=0.93\textwidth]{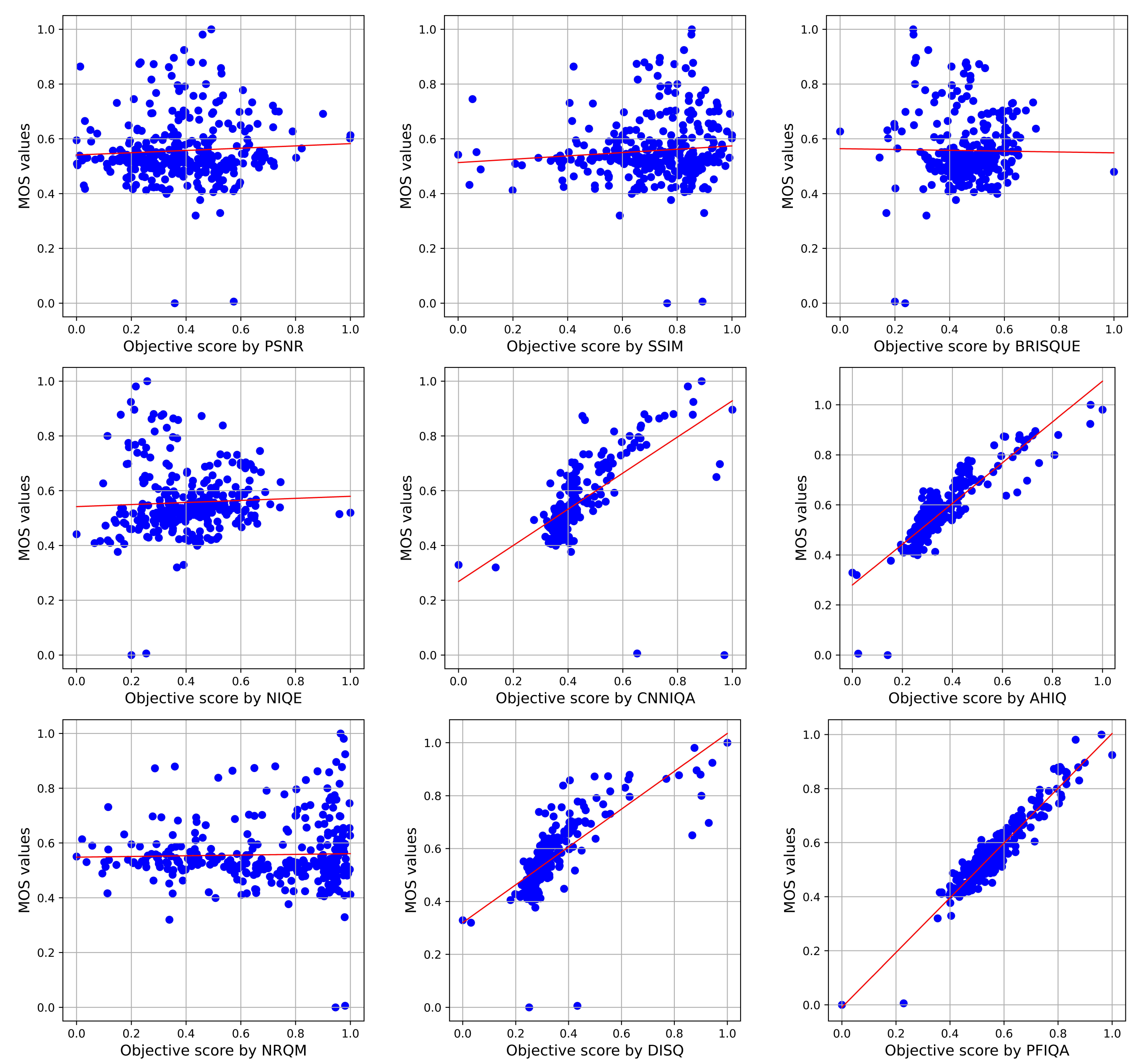}
\caption{Scatter plots of ground-truth mean opinion scores (MOSs) against predicted scores of six generic IQA methods, two SR-IQA methods  and proposed PFIQA on RealSRQ\cite{jiang2022single} dataset. The MOS values have been scaled to range between 0 and 1. Blue points represent the results of the corresponding methods, and the linear fitting of all points is marked by a red straight line. }
\vspace{-4mm}
\label{Fig:curve}
\end{figure*}

\textbf{Quantitative Comparison.}
The primary experimental results are reported in Table \ref{tab:1}, from which we can observe that: 
\textbf{(1)} On the whole, SR-IQA methods outperform most generic IQA methods because there are specifically designed for SR, including the model architecture and training data. However, these learning-based SR-IQA methods employ CNN for feature extraction, which can only focus on local visual information. AHIQ\cite{lao2022attentions} and MANIQA\cite{2022maniqa}, as transformer-based method, enable to capture global visual features, providing a more precise and comprehensive evaluation of image quality, thus outperforming the majority of SR-IQA methods. However, AHIQ has the limitation of requiring HR images for reference. Our PFIQA combines merits of the global modeling of ViT and the local modeling of CNN, along with the prior information of SR, thereby achieving SOTA performance.
\textbf{(2)} Our proposed PFIQA surpasses the compared SR-IQA methods on three benchmarks. Other SR-IQA methods either rely on HR images\cite{zhou2019visual} or fail to utilize SR priors effectively\cite{li2022c,zhou2022quality,fu2023scale}. Even when incorporating LR information, they often overlook the consistency between LR and SR images, leading to suboptimal performance. Our PFIQA not only considers the perceptual quality of SR images but also ensures the consistency between SR and LR images, i.e., reconstruction fidelity. Moreover, by introducing a scale factor, we further guarantee this consistency, achieving SOTA results.
\textbf{(3)} When evaluated on the real-world SR-IQA dataset RealSRQ, PFIQA showcases a substantial improvement compared to other methods. The PLCC and SRCC metrics show an improvement of approximately 5.5$\%$ and 9.07$\%$ compared to the second-best AHIQ and  7.39$\%$ and 10.05$\%$ compared to the third-best MANIQA, which illustrates the superiority of PFIQA in real-world scenarios.

\textbf{Qualitative Analysis.}
Fig. \ref{Fig:score} presents examples of MOS values and predicted scores from DISQ and our PFIQA methods. We can observe that the scores predicted by PFIQA demonstrate better consistency with MOS values compared to DISQ. Furthermore, predicted scores from PFIQA show a strong alignment with the HVS.

To further validate the effectiveness of the proposed PFIQA, Fig. \ref{Fig:curve} presents scatter plots comparing nine metrics with predicted scores through linear fitting. The vertical axis represents subjective MOS values, while the horizontal axis corresponds to predicted scores. Examining the resulting scatter plots, we observe that the points plotted by PFIQA are more compactly distributed and closer to the diagonal line compared to other methods. These impressive results indicate that the quality scores predicted by PFIQA demonstrate a more consistent correlation with subjective assessment of SR images than other comparative methods.

\subsection{Ablation Study}

In this section, we conduct extensive ablation experiments to investigate the effects of various factors on PFIQA for SR-IQA. These experiments include utilizing different reference information, feature extraction and fusion strategies, and fine-tuning techniques. All experiments in this section are performed on the RealSRQ \cite{jiang2022single} dataset.

\textbf{Effects of Reference Information.}
To further investigate the impact of reference information for SR-IQA, we compare the performance of PFIQA utilizing different reference information or not.
In Table \ref{tab:branch}, we can see that: \textbf{(1)} Using only the PA Branch (Table \ref{tab:branch} (a)) can be viewed as a no-reference SR-IQA method, focusing solely on the perceptual quality of the SR image itself, hence leading to sub-optimal performance. \textbf{(2)} Considering the LR image as reference and using only the FA Branch (Table \ref{tab:branch} (b)) can be viewed as a plain reduced-reference SR-IQA method, which accounts for the consistency between the SR and LR images, thereby ensuring reconstruction fidelity to a certain extent. Compared to Table \ref{tab:branch} (a), this brings a significant performance improvement. \textbf{(3)} Combining the PA Branch and FA Branch (Table \ref{tab:branch} (c)), integrates both perceptual quality and reconstruction fidelity, thus obtaining better performance compared to Table \ref{tab:branch} (a,b). \textbf{(4)} Building upon considering the LR image as the reference, and additionally taking into account the scale factor (Table \ref{tab:branch} (d)), PFIQA maximally leverages the available information from SR algorithms, thereby achieving the best SR-IQA performance.
\begin{table}[t]
\centering
\caption{Comparison of leveraging available information of SR algorithms on RealSRQ\cite{jiang2022single}. }

\setlength{\tabcolsep}{5.2pt}
\begin{tabular}{cccc|cc}
\toprule[1.2pt]
\multirow{2}{*}{\#} & \multicolumn{2}{c}{Assessment Branch} & \multirow{2}{*}{Scale Factor} & \multirow{2}{*}{PLCC} & \multirow{2}{*}{SRCC}  \\ 
\cline{2-3}
 & Perception & Fidelity &  &  & \\ 
\midrule
(a) & \checkmark &  &  & 0.9007 & 0.8070 \\ 
(b) &  & \checkmark &  & 0.9233 & 0.8564 \\
\midrule
(c) & \checkmark & \checkmark &  & 0.9242 & 0.8573 \\
(d) & \checkmark & \checkmark & \checkmark & \textbf{0.9269} & \textbf{0.8597}  \\
\bottomrule[1.2pt]
\end{tabular}
\vspace{-4mm}
\label{tab:branch}
\end{table}

\begin{table}[t]
\centering
\caption{Comparison of different visual feature extraction and fusion strategies on RealSRQ\cite{jiang2022single}. }

\begin{tabular}{cccc|cc}
\toprule[1.2pt]
\multirow{2}{*}{\#} & \multicolumn{2}{c}{Feature} & \multirow{2}{*}{Fusion Method} & \multirow{2}{*}{PLCC} & \multirow{2}{*}{SRCC}  \\ 
\cline{2-3}
 & ResNet & ViT &  &  & \\
 \midrule
(a) & \checkmark &  & - & 0.9070 & 0.8302 \\ 
(b) &  & \checkmark & - & 0.9010 & 0.8105 \\
\midrule
(c) & \checkmark & \checkmark & Concatenation& 0.9198 & 0.8460 \\
(d) & \checkmark & \checkmark & Adaptive Fusion & \textbf{0.9269} & \textbf{0.8597}  \\
\bottomrule[1.2pt]
\end{tabular}
\vspace{-4mm}
\label{tab:fusion}
\end{table}

\begin{table}[t]
\centering
\caption{Comparison of using different different fine-tuning strategies on RealSRQ\cite{jiang2022single}. }

\setlength{\tabcolsep}{5.2pt}
\begin{tabular}{m{0.9cm}<{\centering}m{1.5cm}<{\centering}m{1.50cm}<{\centering}|m{1.5cm}<{\centering}m{1.5cm}<{\centering}}
\toprule[1.2pt]
\multirow{2}{*}{\#} & \multicolumn{1}{c}{ViT} &\multicolumn{1}{c|}{ResNet} & \multirow{2}{*}{PLCC} & \multirow{2}{*}{SRCC}  \\ 
\cmidrule{2-3}
 & Updated  & Updated   &  & \\ 
\midrule
(a) &&\checkmark& 0.9025 & 0.8277 \\ 
(b) & \checkmark  && 0.9195 & 0.8539 \\
(c) &\checkmark&\checkmark& 0.9065 & 0.8305 \\

\midrule
(d) && & \textbf{0.9269} & \textbf{0.8597}  \\
\bottomrule[1.2pt]
\end{tabular}
\vspace{-3mm}
\label{tab:update}
\end{table}

\textbf{Effects of Different Feature Extraction and Fusion Strategies.}
Since we use ViT and ResNet to extract global and local visual features from images, it is necessary to investigate how they impact the performance of SR-IQA.
In Table \ref{tab:fusion}, we can find that: \textbf{(1)} Using only each of the visual features extracted by ViT and ResNet (Table \ref{tab:fusion} (a,b)) can achieve satisfactory SR-IQA results compared to other methods in Table \ref{tab:1}. 
\textbf{(2)} Utilizing both global and local visual features (Table \ref{tab:fusion} (c)), leads to better performance compared to Table \ref{tab:fusion} (a,b). This demonstrates the complementary nature of global and local features for comprehensive SR-IQA.
\textbf{(3)} Further analysis of the adaptive fusion of global and local features, as indicated in Table \ref{tab:fusion} (d), reveals that PFIQA attains superior performance. This demonstrates the efficacy of the AFMs in learning the importance of features from global and local features.

\textbf{Effects of Different Fine-tuning Strategies.}
To investigate the impact of fine-tuning ViT and ResNet on the performance of SR-IQA, we conducted experiments with different parameter updating strategies.
In Table \ref{tab:update}, we observe that: \textbf{(1)} Keeping both pre-trained ViT and ResNet parameters fixed (Table \ref{tab:update} (d)) yields the best performance, with PLCC of 0.9269 and SRCC of 0.8597. This suggests that the pre-trained knowledge from these models is highly valuable for SR-IQA task. This approach helps maintain the model's generalization capability and mitigates the risk of overfitting on small, specialized datasets like those used in SR-IQA tasks.
\textbf{(2)} Updating either ViT or ResNet individually (Table \ref{tab:update} (a,b)) leads to a decrease in performance compared to the non-updated version (Table \ref{tab:update} (d)). This indicates that fine-tuning on the small SR-IQA dataset may lead to overfitting and loss of generalized features learned during pre-training.
\textbf{(3)} Updating both ViT and ResNet (Table \ref{tab:update} (c)) results in further performance degradation. This reinforces the observation that preserving the pre-trained knowledge is crucial for the tsak of SR-IQA task, especially given the limited scale of the SR-IQA datasets.

\section{Conclusion}
In this paper, we propose a novel reduced-reference Perception- and Fidelity-aware SR-IQA (PFIQA) network, which integrates LR images and the scale factors as prior knowledge to comprehensively assess the perceptual quality and reconstruction fidelity of SR images. We leverage the merits of global modeling of ViT and local relation of CNN to enable comprehensive visual feature extraction, and incorporate the scale factors to obtain features that are more relevant to SR-IQA.
The extensive results from three  benchmark datasets showcase the efficacy of PFIQA, and the evaluation results are highly consistent with HVS.

\bibliographystyle{IEEEtran}
\bibliography{ref}

\end{document}